# On the propagation of sound waves in a stellar wind traversed by periodic strong shocks

**F. P. Pijpers**

Uppsala Astronomical Observatory, Box 515, S-75120 Uppsala, Sweden

**Abstract.** It has been claimed that in stellar winds traversed by strong shocks the mechanism for driving the wind by sound wave pressure cannot operate because sound waves cannot propagate past the shocks. It is shown here that sound waves can propagate through shocks in one direction and that this is a sufficient condition for the sound wave pressure mechanism to work. A strong shock amplifies a sound wave passing through it and can drag the sound wave away from the star. It is immaterial for the sound wave pressure gradient that the sound wave vector points towards the star. Since the strong shocks drag the sound waves away, the star itself is the source for the sound waves propagating towards it.

**Key words:** Shock waves – Stars : oscillations – Stars : mass loss

## 1. Introduction

The effect of the presence of sound waves on the winds from stars can be modelled by introducing a wave pressure (Pijpers & Hearn, 1989). The gradient of this pressure acts in the same way that the gradient of the gas pressure does. For a linear sound wave which has a small amplitude the pressure can be calculated explicitly without a knowledge of the background velocity as long as this is smooth. The appropriate expressions are derived by Pijpers and Hearn (1989) on the basis of the linear perturbation formulations for the propagation of sound waves in continuous moving media of Bretherton (1970).

The linear theory of sound wave driven winds has been used to model the wind of cool giants (Pijpers and Habing, 1989). It has also been used to model to outflow from the M supergiant VX Sgr (Pijpers, 1990). Objections have been raised (e.g. Wood, 1990) to the application of this model to the Mira class of variable M giants. The reasoning is that the stellar pulsation of Mira variables generates strong shocks which propagate out into the wind. These strong shocks would overtake and absorb the sound waves so that the pressure from sound waves would be effectively cancelled. The purpose of this paper is to demonstrate that this is incorrect.

The expression for the wave pressure tensor in the linear regime is :

$$\mathbf{P}^{\rm l}_{{\rm w}ij} = -\frac{1}{2}c_{\rm S}^2\widetilde{\rho}a^2|\mathbf{k}|^2\left(\frac{k_i k_j}{|\mathbf{k}|^2} + \frac{1}{2}(\gamma-1)\delta_{ij}\right) \quad (1)$$

Here $\mathbf{k}$ is the wave vector of the sound wave and $\delta_{ij}$ is the Kronecker delta. $c_{\rm S}$ is the speed of sound defined by

$$c_{\rm S}^2 = \frac{\partial p}{\partial \rho} = \gamma\frac{p}{\rho} \quad (2)$$

where $p$ is the ordinary gas pressure and $\rho$ the gas density. $\widetilde{\rho}$ is the Lagrangian mean density of the flow. $a|\mathbf{k}|$ is the amplitude of the displacement in the sound wave. $\gamma$ is the adiabatic index. If the waves become non-linear the approximations of this linear perturbation method are no longer valid.

It is still possible to derive a pressure tensor for non linear waves as long as the velocity amplitude of the sound wave is less than the sound speed. To do this one can follow the generalized Lagrangian mean formalism (GLM) presented by Andrews & MacIntyre (1978a, 1978b). Koninx & Pijpers (1992) show that if the disturbance profile of the sound waves is written as :

$$\xi = a\frac{\mathbf{k}}{|\mathbf{k}|}f(\mathbf{k}\cdot\mathbf{x} - \omega t - \theta) \quad (3)$$

the general wave pressure tensor takes the form :

$$\mathbf{P}^{\rm nl}_{{\rm w}ij} = \frac{1}{2\pi}\int_0^{2\pi} p\left(\delta_{ij} - \frac{\delta_{ij} + \left[\delta_{ij} - \frac{k_i k_j}{|\mathbf{k}|^2}\right]a|\mathbf{k}|f'}{(1+a|\mathbf{k}|f')^\gamma}\right)d\theta \quad (4)$$

$f'$ is the derivative of the profile function $f$. The limitation to the velocity amplitude of the sound wave that is less than the sound speed means that $a|\mathbf{k}| < 1$.

Note that the profile function $f$ gives considerable freedom of the shape of the sound wave profile. The waves are not required to be sinusoidal. It is well known that sound waves in stellar atmospheres very quickly deform into a saw-tooth shape with small discontinuities. As long as the velocity amplitude of the saw-tooth is less than the sound speed the conditions are met for the non-linear wave pressure formalism to be valid. For an isothermal saw-tooth wave this means that the pressure ratio at each discontinuity should be less than $3/2 + 1/2\sqrt{5} \approx 2.6$. For adiabatic saw-tooth waves in a gas with a polytropic equation of state the pressure jump should be less than $1 + 1/4\gamma(\gamma + 1) + 1/4\sqrt{(\gamma + 1)^2 + 16}$. This is approximately 3.31 if the polytropic index $\gamma = 5/3$.

## 2. The propagation of sound waves

The propagation of sound waves in an outflow traversed by periodic strong shocks can be described most conveniently in two steps. The first step concerns the behaviour of the wave at the discontinuity. The second concerns the propagation through the continuous part of the wind between the shock discontinuities.

In the coordinate frame where the strong shock wave generated by the pulsation is at rest the inflow into the shock is supersonic and the outflow is subsonic as long as $(\partial^2 V/\partial p^2)_{\rm S} > 0$ where $V \equiv 1/\rho$ is the specific volume of the gas (cf. Landau & Lifshitz, 1959).

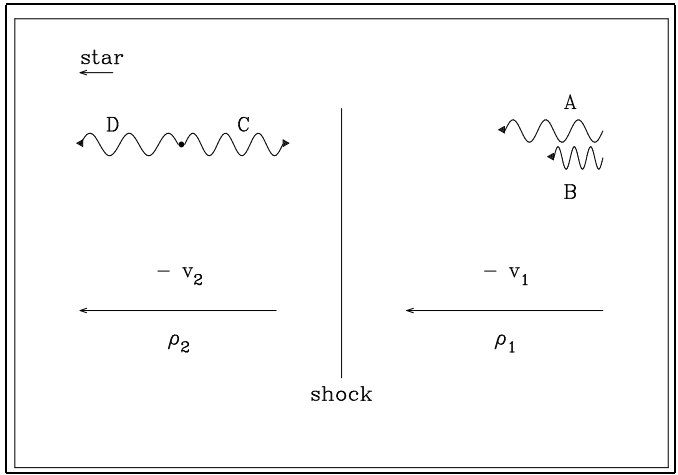

**Fig. 1.** The velocities and densities in the coordinate frame in which the shock is at rest. Note the sign definitions so that the values of $v_1$ and $v_2$ are positive

A sound wave with a wave vector pointing away from the discontinuity in the upwind direction (case B in fig. 1) will therefore be overtaken by the shock. Of course a sound wave with a wave vector pointing towards the discontinuity from the upwind direction (case A) will reach the shock as well. Both types of sound waves will pass through the shock and can then propagate away from it with the flow in the downwind direction. The amplitude of the sound waves will be increased after passage through the shock which is shown in the next section. Sound waves can also propagate towards the discontinuity in the downwind part of the flow (case C of fig. 1) since this flow is subsonic. Upon reaching the discontinuity these waves must relinquish their energy to the shock wave or be reflected since they cannot propagate into a flow that is supersonic in the direction opposite to their wave vector. It can be shown that such waves are partially reflected (Landau & Lifshitz, 1987). Naively the cases B and C are the ones encountered when shocks and sound waves both propagate away from the star.

It is important at this point to realize that the analysis presented above is done in the frame comoving with the shock. The frame in which the star is at rest is the coordinate frame in which the equations for the wind are derived. Therefore it is important to know the direction of propagation in this coordinate frame. It is clear from the above that after the passage of just one shock through a given parcel of gas only the sound waves of type A (or equivalently D) survive. This does give waves of type B (or equivalently C) some time to travel partway out from the zone where the sound is generated. Determination of the maximum radius at which waves of type B/C can occur is not trivial. The sound waves are most likely to be generated at the top of the convectively unstable layer at radius $r_{\rm c}$. If $r_{\rm S}$ is the distance the strong shock wave travels in a time $t$ after passing $r_{\rm c}$ then it satisfies:

$$\frac{\partial r_{\rm S}}{\partial t} = v_{\rm S}(r_{\rm S}, t)$$
$$r_{\rm S} = r_{\rm c} \qquad \text{at } t = 0 \tag{5}$$

Here $v_{\rm S}$ is the velocity of the shock wave in the rest frame of the star. For the sound waves of type B/C emitted from the top of the convective layer at some phase delay $\alpha$ after the passage of a shock, the distance a wave front can travel is:

$$\frac{\partial r_{\rm sw}}{\partial t} = v(r_{\rm sw}, t) + c_{\rm S}(r_{\rm sw}, t)$$
$$r_{\rm sw} = r_{\rm c} \qquad \text{at } t = -\frac{\alpha}{2\pi}P \tag{6}$$

Here $v$ is the velocity of the fluid flow in the rest frame of the star, $c_{\rm S}$ the sound speed, and $P$ the period between passages of the strong shock waves. The sound waves are overtaken by the next shock when $r_{\rm sw} = r_{\rm S} \equiv r_\alpha$. The maximum distance that sound waves can travel out from the star is then $r_{\rm f} \equiv \max r_\alpha$ as a function of $\alpha$. Although it is possible in principle that $r_{\rm f}$ is infinite, it is likely that beyond some finite radius all the gas will have passed at least once through a shock front. The region where waves of type A/D are generated, either directly from turbulent convection or through shock processing of waves of type B/C, extends up to this radius $r_{\rm f}$. Beyond that radius only waves of type A/D survive. In the rest of this section therefore only the case A of figure 1 is considered.

In the frame in which the star is at rest the velocity of the flow emerging from the discontinuity at time $t = 0$ at a fixed position with respect to the star is:

$$v(t) = v_{\rm S} - v_2 + (v_2 - v_1)\frac{t}{P} \tag{7}$$

The signs have been chosen such that all velocities on the right hand side in equation (7) have positive values. On the left hand side $v(t)$ is positive for material moving away from the star. It is assumed that the velocity is linear as a function of distance behind the shock and therefore also as a function of time. This linearity is a very good approximation for strong shocks (cf. Landau & Lifshitz, 1959). The hydrodynamical simulations of Bowen (1988) show this behaviour for the velocity also. The average velocity in the Eulerian reference frame of a sound wave in this period $P$ on emerging from the shock is :

$$V_{\rm sw} = \frac{1}{P} \int_0^P v(t) - c_{\rm S}(t) \, {\rm d}t \qquad (8)$$

If $V_{\rm sw} < 0$ a sound wave starting out in the wind can move further towards the star than it gets dragged outwards by the mean flow between the strong shocks. In this case the passage of one of the strong shocks absorbs the sound waves moving in the same direction as the shock and the sound waves propagating in the other direction can 'drain' back into the star. If $V_{\rm sw} > 0$ a sound wave can only travel part of the way towards the next discontinuity and it is dragged away from the star by the mean flow. Even if a sound wave starts out in the stellar photosphere with a wave vector pointing towards the star the flow between the strong shocks will drag it out if $V_{\rm sw} > 0$. Combining the equations above yields :

$$V_{\rm sw} = c_{\rm S1} \frac{1}{2\pi} \int_0^{2\pi} \left[ M(\phi) - \frac{c_{\rm S}(\phi)}{c_{\rm S1}} \right] {\rm d}\phi \qquad (9)$$

Here $\phi \equiv 2\pi t/P$ is the phase of the pulsation cycle. All velocities are now dimensionless Mach numbers since they are divided by the sound speed. The shock speed is divided by the sound speed in front of the shock $c_{\rm S1}$. The Mach number of the flow $M(\phi)$ is :

$$M(\phi) = M_{\rm S} - \frac{M_1}{S} - M_1 \left(1 - \frac{1}{S}\right) \frac{\phi}{2\pi} \qquad (10)$$

Here $S$ is the ratio of densities behind and in front of the shock. Making use of the continuity equation at the discontinuity there is a relation between $v_1$ and $v_2$ :

$$\frac{\rho_2}{\rho_1} = \frac{v_1}{v_2} = \frac{(\gamma+1)M_1^2}{(\gamma-1)M_1^2 + 2} \equiv S \qquad (11)$$

which holds if the shock waves are adiabatic. If the shocks radiate, which is more realistic, the law of conservation of mass still holds but $S$ is larger.

In Miras the shocks in the outer atmosphere and wind are radiative. The temperature variations are largest for adiabatic waves in which case :

$$\frac{T_2}{T_1} = \frac{\left(2\gamma M_1^2 - (\gamma-1)\right)\left((\gamma-1)M_1^2 + 2\right)}{(\gamma+1)^2 M_1^2} \qquad (12)$$

For radiative shocks the temperature ratio is smaller. In the limiting case of isothermal shocks the ratio is of course unity. For isothermal shocks and a flow that is also isothermal the ratio $c_{\rm S}(\phi)/c_{\rm S1} = 1$. For adiabatic shocks and flow the ratio can be calculated from the equation of conservation of energy and (10) :

$$\left(\frac{c_{\rm S}(\phi)}{c_{\rm S1}}\right)^2 = \frac{1}{2}(\gamma-1)\left[M_1^2 - (M(\phi) - M_{\rm S})^2\right] + 1 \qquad (13)$$

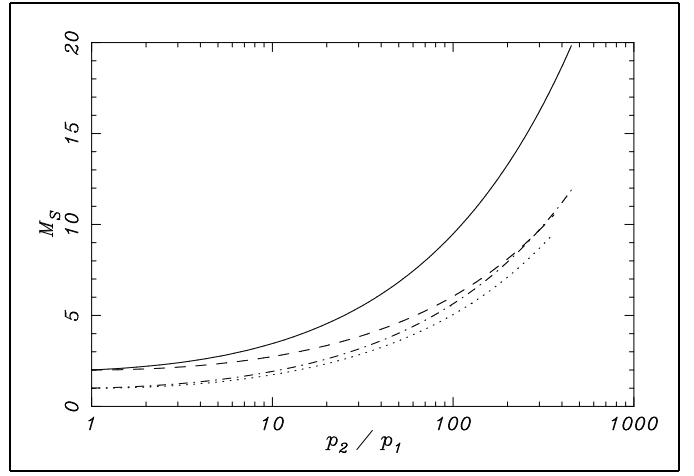

**Fig. 2.** The minimum Mach number of the shock $M_{\rm S}$ for which sound waves are dragged out into the wind. The full line is for a flow with adiabatic shocks, the dashed line for a flow with isothermal shocks. The dash-dotted line is the minimum $M_S$ for which the average fluid velocity is outward for a flow with adiabatic shocks. The dotted line is the same for a flow with isothermal shocks

The condition for $V_{\rm sw}$ leads to a lower limit for $M_{\rm S}$ so that sound waves cannot drain back to the star. This limit can be calculated by carrying out the integration in (9) using (10) and also the equation of conservation of energy to determine $c_{\rm S}(\phi)$.

The result is shown in figure 2. Also shown in 2 is the minimum shock Mach number for which the average fluid flow is outward. If $M_{\rm S}$ is less than the values shown with the dash-dotted curve, for adiabatic flow, and the dotted curve, for isothermal flow, the average fluid flow is towards the star which describes accretion instead of mass loss. There is therefore only a small range of shock speeds for which the sound waves can 'drain' back into the star and this range becomes smaller as the shocks radiate more.

This proves that waves propagating 'towards' the star in the frame in which the shock is stationary can actually be dragged away from the star by the mean flow. Therefore in general sound waves do propagate away from the star albeit 'backwards'. As a consequence the wave energy density is not equal to 0 in the stellar wind.

As an example one can use the results of Bowen (1988) to estimate the shock strength at around 2 stellar radii and compare the minimum required shock speed with the actual speed of the shock. In Bowen's model without dust (his fig. 5) the shock strength in terms of a pressure ratio is around 15. The propagation speed of the shock is around 15 km/s which is close to the minimum speed for isothermal shocks above which sound waves are dragged out by the flow. This makes it very hard to conclude whether or not the simulations yield $V_{\rm sw} > 0$. In Bowen's figure 5 the shocks are clearly not isothermal but the radiative cooling law was necessarily primitive. A more complete treatment of the cooling would probably make the shocks more radiative and therefore more closely isothermal. This would also reduce the pressure ratio at the shock. Both ef-fects would put the simulations more firmly into the region of fig. 2 where sound waves are dragged out into the wind.

It is important to note that both in the linear and in the non-linear wave pressure tensor (eqs. (1) and (4)) the sign of the wave vector cancels out if it is assumed to be purely in the radial direction. It is therefore immaterial that the wavevector points in the direction of the star in the comoving frame. The only requirement is that the wave energy density is not equal to 0 and this requirement is satisfied.

Summarizing : up to a point $r_\mathrm{f}$ in the wind waves of all types occur and waves of type A/D are generated. Beyond this point only waves of type A/D can survive. The strong shock waves generated by the pulsation can drag sound waves of type A/D superposed on the flow with them so that these will move out into the wind. The condition for this in essence is that the average over a full pulsation cycle of the outflow velocity is larger than the average over a full pulsation cycle of the sound speed. The results of numerical simulations to date (e.g. Wood, 1979 ; Bowen, 1988) do allow this to occur. It is immaterial for the wave pressure that only those sound waves with a wavevector pointing towards the star in the fluid frame (type A/D) can survive the passage of the strong shock waves.

## 3. The amplification of sound waves

On the passage through a shock the amplitude of the sound wave increases. The proof of this is readily available (Landau & Lifshitz, 1959) but is repeated here for convenience. The starting point is the solution of the equations of conservation of mass, of conservation of momentum, and of conservation of energy, where the plane sound wave is assumed to impinge on the shock along the normal. The coordinate frame is chosen to comove with the shock (see fig. 1). These equations are :

$$
\begin{aligned}
\rho_1 v_1 &= \rho_2 v_2 \\
p_1 + \rho_1 v_1^2 &= p_2 + \rho_2 v_2^2 \\
w_1 + \frac{1}{2} v_1^2 &= w_2 + \frac{1}{2} v_2^2
\end{aligned} \tag{14}
$$

If the velocity amplitude of the sound wave impinging on the shock is $\delta v_1$ the shock wave will start to oscillate with an amplitude $\delta u$ and the equations above must be modified. In a linear perturbation analysis the equations become

$$
\begin{aligned}
v_1 \delta \rho_1 + \rho_1(\delta v_1 - \delta u) &= v_2 \delta \rho_2 + \rho_2(\delta v_2 - \delta u) \\
\delta p_1 + v_1^2 \delta \rho_1 + 2\rho_1 v_1(\delta v_1 - \delta u) &= \\
&\quad \delta p_2 + v_2^2 \delta \rho_2 + 2\rho_2 v_2(\delta v_2 - \delta u) \\
\delta w_1 + v_1(\delta v_1 - \delta u) &= \delta w_2 + v_2(\delta v_2 - \delta u)
\end{aligned} \tag{15}
$$

The incident sound wave is assumed to be adiabatic :

$$
\begin{aligned}
\delta s_1 &= 0 \\
\delta v_1 &= \frac{c_{S1}}{\rho_1}\delta \rho_1 = \frac{\delta p_1}{c_{S1}\rho_1} \\
\delta w_1 &= \frac{\delta p_1}{\rho_1}
\end{aligned} \tag{16}
$$

Just behind the shock there are two types of disturbances. There is an entropy wave which is stationary with respect to the fluid. There is also a sound wave. The disturbances behind the shock are the sum of the two where ($S$) refers to the sound wave and ($e$) refers to the entropy wave :

$$
\begin{aligned}
\delta \rho_2 &= \delta \rho_2^{(S)} + \delta \rho_2^{(e)} \\
\delta v_2 &= \delta v_2^{(S)} + \delta v_2^{(e)} \\
\delta p_2 &= \delta p_2^{(S)} + \delta p_2^{(e)} \\
\delta w_2 &= \delta w_2^{(S)} + \delta w_2^{(e)}
\end{aligned} \tag{17}
$$

The relations between the quantities in the sound wave are :

$$
\begin{aligned}
\delta s_2^{(S)} &= 0 \\
\delta v_2^{(S)} &= \frac{c_{S2}}{\rho_2}\delta \rho_2^{(S)} = \frac{\delta p_2^{(S)}}{c_{S2}\rho_2} \\
\delta w_2^{(S)} &= \frac{\delta p_2^{(S)}}{\rho_2}
\end{aligned} \tag{18}
$$

and for the entropy wave the following holds :

$$
\begin{aligned}
\delta p_2^{(e)} &= 0 \\
\delta v_2^{(e)} &= 0 \\
\delta w_2^{(e)} &= T_2 \delta s_2^{(e)} = -c_{S2}^2 \frac{\delta \rho_2^{(e)}}{(\gamma - 1)\rho_2}
\end{aligned} \tag{19}
$$

The ratio of the amplitudes of the pressure perturbation of the sound waves $\delta p_2^{(S)}/\delta p_1$ can be determined from the set of linear equations (15). It is :

$$
\begin{aligned}
\left(\frac{\delta p_2^{(S)}}{\delta p_1}\right)_A &= \frac{2M_1 M_2^2(M_1^2 - 1) - (1 + M_1)[M_1^2 + \frac{2}{(\gamma-1)}]}{2M_2^2(M_1^2 - 1) - (1 + M_2)[M_1^2 + \frac{2}{(\gamma-1)}]} \\
&\quad \times \frac{M_1 + 1}{M_2 + 1} \\
&= \frac{(1 + M_1)^2}{1 + 2M_2 + 1/M_1^2}\left[1 - \frac{\gamma - 1}{\gamma + 1}\left(1 - \frac{1}{M_1}\right)^2\right]
\end{aligned} \tag{20}
$$

A more general and somewhat more lengthy derivation is given in the second edition of 'Fluid Mechanics' of Landau & Lifshitz (1987). The velocity perturbation is :

$$
\frac{\delta v_2^{(S)}}{\delta v_1} = \left[\frac{2 + (\gamma - 1)M_1^2}{2\gamma M_1^2 - (\gamma - 1)}\right]^{1/2} \frac{1}{M_1} \frac{\delta p_2^{(S)}}{\delta p_1} \tag{21}
$$

For the purposes of sound wave driven winds the quantity of interest is the wave action density $\mathcal{A}$ which is conserved in the absence of wave excitation and dissipation. For small amplitude sound waves it is :

$$
\mathcal{A} \equiv \frac{E_\mathrm{w}}{\widehat{\omega}} \tag{22}
$$

Here $E_\mathrm{w}$ is the wave energy and $\widehat{\omega} \equiv |\omega - \mathbf{u} \cdot \mathbf{k}|$ is the intrinsic wave frequency. For waves propagating in the same direction as the flow this reduces to :

$$
\mathcal{A} \equiv \frac{1}{2}\rho \delta v^2 \frac{(v + c_\mathrm{S})}{\omega c_\mathrm{S}} \tag{23}
$$

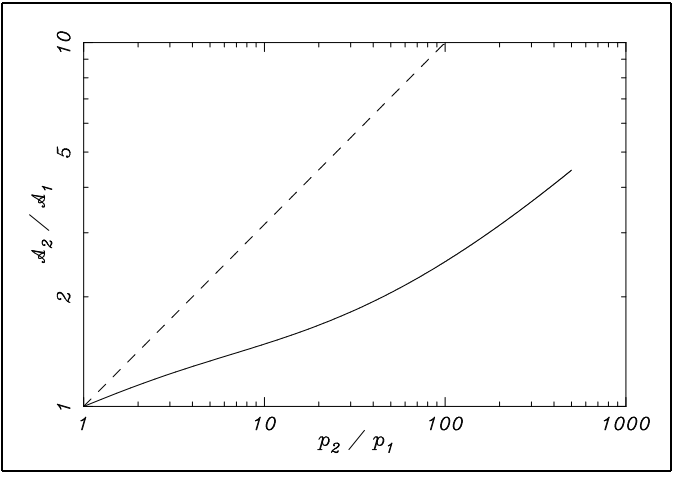

**Fig. 3.** The ratio of the wave action density in front of and behind the shock as a function of shock strength, for case A of fig. 1. The full line is for an adiabatic shock, the dashed line for an isothermal shock

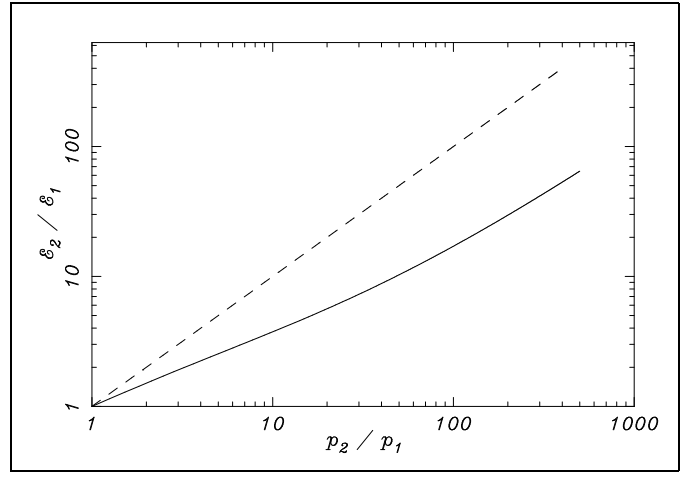

**Fig. 4.** The ratio of the wave energy density behind and in front of the shock as a function of shock strength, for case A of fig. 1. The full line is for an adiabatic shock, the dashed line for an isothermal shock

Here the factor $1/2$ is derived assuming a sinusoidal wave profile. For a saw-tooth profile of the sound waves this factor should be $1/3$. Using equation (21) the ratio of the wave action density in front of the shock and behind the shock can easily be calculated :

$$\left(\frac{\mathcal{A}_2}{\mathcal{A}_1}\right)_A = \frac{(\gamma+1)}{2\gamma M_1^2 - (\gamma-1)} \frac{1+M_2}{1+M_1} \left(\frac{\delta p_2^{(S)}}{\delta p_1}\right)_A^2 \qquad (24)$$

Equation (24) is the wave action density transmission coefficient appropriate for waves of type A of figure 1 passing through an adiabatic shock.

For a strongly radiative shock the perturbation of the radiative term in the energy equation must also be taken into account. If it is assumed that this term is negligible for (nearly) isothermal shocks the ratio of the wave action densities can easily be calculated. The pressure perturbation ratio becomes :

$$\left(\frac{\delta p_2^{(S)}}{\delta p_1}\right)_A = M_1^2(1+M_1)\frac{[M_1 + \frac{(\gamma-1)}{(2-\gamma)}]}{[M_1^2 + \frac{M_1}{(2-\gamma)} + \frac{(\gamma-1)}{(2-\gamma)}]} \qquad (25)$$

The wave action density ratio becomes :

$$\left(\frac{\mathcal{A}_2}{\mathcal{A}_1}\right)_A = \frac{1}{M_1^3}\left(\frac{\delta p_2^{(S)}}{\delta p_1}\right)_A^2 \qquad (26)$$

The reflection coefficient is zero for both adiabatic shocks and isothermal shocks.

In figure 3 the transmission coefficient as a function of shock strength is shown for an adiabatic shock and an isothermal shock. The value of $\gamma$ is taken $5/3$ in both cases. This means that **away from the shock** the gas is assumed always to behave as an ideal mono-atomic gas. It is clear that if there is no pressure discontinuity nothing happens to the sound waves and the ratio of the wave action density behind and in front of the shock front is unity. Because the density ratio and the velocity ratio are higher for isothermal shocks than for adiabatic shocks, isothermal shocks are more efficient at amplifying sound waves.

The wave action density is the proper quantity to measure the effect of shocks on sound waves since the wave action density is conserved in the absence of mechanisms generating or dissipating waves. However, it is easier to interpret the effect of shocks in terms of the energy density in the sound waves. This can easily be calculated from the definition of the wave action density for linear sound waves (22). The result for the waves of type A is shown in figure 4.

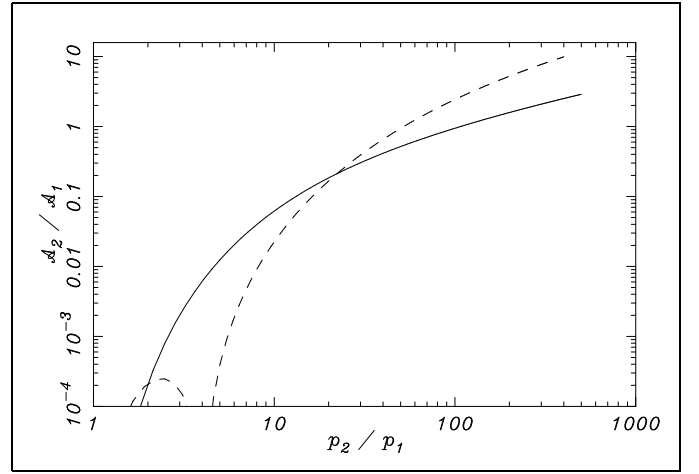

**Fig. 5.** The ratio of the wave action density behind and in front of the shock as a function of shock strength, for case B of fig. 1. The full line is for an adiabatic shock, the dashed line for an isothermal shock

For waves of type B the reflection coefficient is also equal to 0. The transmission coefficient for type B waves being overtaken by the shock differs slightly from the one for type A waves. For transition through an adiabatic shock the pressure perturbation ratio is given by :

$$\left(\frac{\delta p_2^{(S)}}{\delta p_1}\right)_B = \frac{(1-M_1)^2}{1+2M_2+1/M_1^2} \quad (27)$$
$$\times \left[1 - \frac{\gamma-1}{\gamma+1}\left(1+\frac{1}{M_1}\right)^2\right]$$

and the ratio of wave action densities by :

$$\left(\frac{\mathcal{A}_2}{\mathcal{A}_1}\right)_B = \frac{(\gamma+1)}{2\gamma M_1^2 - (\gamma-1)}\frac{M_2+1}{M_1-1}\left(\frac{\delta p_2^{(S)}}{\delta p_1}\right)_B^2 \quad (28)$$

For transition through an isothermal shock the pressure perturbation ratio is given by :

$$\left(\frac{\delta p_2^{(S)}}{\delta p_1}\right)_B = M_1^2 \frac{(1-M_1)^2}{(1+M_1)}\frac{[M_1 - \frac{(\gamma-1)}{(2-\gamma)}]}{[M_1^2 + \frac{M_1}{(2-\gamma)} + \frac{(\gamma-1)}{(2-\gamma)}]} \quad (29)$$

and the ratio of wave action densities by :

$$\left(\frac{\mathcal{A}_2}{\mathcal{A}_1}\right)_B = \frac{1}{M_1^3}\frac{M_1+1}{M_1-1}\left(\frac{\delta p_2^{(S)}}{\delta p_1}\right)_B^2 \quad (30)$$

The transmission coefficient as a function of shock strength is shown in figure 5. It is clear that in the limit where there is no discontinuity and the ratio of the pressure approaches unity it should be impossible to convert waves of type B into waves of type D. The transmission coefficient indeed decreases to 0 in this limit. For very strong shocks the Mach number of the material flowing into the shock is very high. It should therefore matter little whether a sound wave is of type A or type B since the propagation velocity with respect to the shock is almost the same. Comparing figures 3 and 5 does indeed show that the transmission coefficients for the two wave types are of the same order of magnitude. For isothermal shocks there is also a zero transmission coefficient for waves of type B for :

$$\frac{p_2}{p_1} = \left(\frac{\gamma-1}{2-\gamma}\right)^2 \quad (31)$$

It is possible that this is due ignoring the perturbation of the radiative term in the energy equation at the shock for sound waves passing an isothermal shock. If the shock front is radiative and the presence of sound waves is properly taken into account in the radiation term the ratio of the wave action densities may remain finite for all shock strengths larger than unity.

For waves of type C the transmission coefficient is zero and the reflection coefficient can be shown to be (Landau & Lifshitz, 1987) :

$$\frac{\mathcal{A}_r}{\mathcal{A}_i} = \frac{1+M_2}{1-M_2}\left(\frac{1-2M_2+1/M_1^2}{1+2M_2+1/M_1^2}\right)^2 \quad (32)$$

This relation holds for reflection at adiabatic shocks and isothermal shocks, but one should keep in mind that the relation between $M_2$ and $M_1$ is different for the two cases.

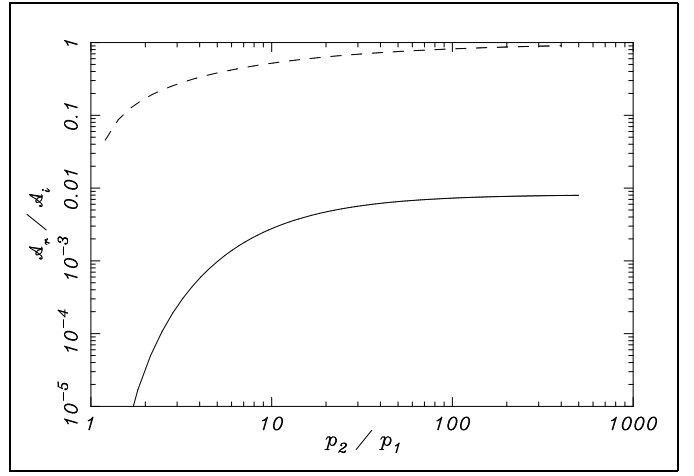

**Fig. 6.** The ratio of the wave action density of reflected and incident waves as a function of shock strength, for case C of fig. 1. The full line is for an adiabatic shock, the dashed line for an isothermal shock

In the limit of a vanishingly small discontinuity the reflection coefficient should become zero since there is no discontinuity to interfere with the sound waves any more. For isothermal shocks the ratio of the densities becomes infinite for infinite shock strength. This must have the consequence that waves are reflected perfectly and therefore the reflection coefficient must approach 1. For adiabatic waves the density ratio remains finite and therefore the reflection coefficient remains less than 1. In figure 6 the reflection coefficient is shown as a function of shock strength. It shows the correct limiting behaviour.

## 4. Implications for winds driven by sound wave pressure

### 4.1. A periodic solution for $\mathcal{A}$

From the previous sections it is clear that the presence of strong shocks in the atmospheres and winds of red giants does not preclude the presence of sound waves. In the modelling of the winds of such stars it is therefore important to take both effects into account. The transport equation for the wave action density of sound waves is (cf Koninx & Pijpers, 1992) :

$$\frac{\partial \widetilde{\rho}\mathcal{A}}{\partial t} + \nabla \cdot [\widetilde{\rho}\mathcal{A}u + \mathcal{B}] = \widetilde{\rho}\mathcal{F} \quad (33)$$

Here $\mathcal{B}$ is a flux term and $\mathcal{F}$ collects all the source terms. For linear sound waves this equation reduces to :

$$\frac{\partial \mathcal{A}}{\partial t} + \nabla \cdot (v_{sw}\mathcal{A}) = \mathcal{A}\ln(T_A)\delta\left(t - (n + \frac{\alpha(r)}{2\pi})P\right) + \\ - \frac{v_{sw}\mathcal{A}}{L} \quad n = 1, 2, 3, ... \quad (34)$$

Here $v_{sw}$ is the (instantaneous) propagation speed of the sound waves in the rest frame of the star. The second term on the right hand side of (34) represents the weak damping of linear sound waves in between the strong shocks. The first term on the right hand side represent the periodic sound wave amplification due to the passage of the shocks. The factor $T_A$ is the ratio of the wave action densities behind the shock and in front of the shock and it can be calculated from (24) for an adiabatic shock and (26) for an isothermal shock. Even for a strictly periodic pulsation cycle the strength of the shocks can be a function of radius, and therefore so is $T_A$. The reduced phase angle $\alpha(r)$ is introduced because the shocks pass two different points $r$ and $r'$ in the wind at different phases of the pulsation cycle. Note that the occasional input of wave action from waves of type B/C below the maximum radius $r_f$ to which they can propagate is ignored in (34). Only waves of type A are taken into consideration. The general transport equation for the wave action density is more complicated to solve so for clarity the linearized equation is used here instead. As was shown by Pijpers and Koninx (1992) this means that the results are a reasonable approximation as long as the velocity amplitude of the waves is less than about half the sound speed.

The only trivial solution of (34) is $\mathcal{A} \equiv 0$. That solution is unstable and the effect of sound waves on a stellar wind traversed by shocks needs to be modelled explicitly using time dependent hydrodynamical simulations with the appropriate sound wave pressure term added to the transport equations of momentum and of energy, and with (34) as an extra equation to solve at every time step. However it is useful to consider substituting a solution for the wave action density of the form :

$$\mathcal{A} = \mathcal{A}_0(t) \frac{|V_{sw0}| r_0^2}{|V_{sw}| r^2} \exp\left[-\frac{(r - r_0)}{L_{\text{eff}}}\right] \quad (35)$$

Here $V_{sw0}$ is the sound wave propagation velocity at radius $r_0$ from the star, averaged over a period according to (8). This expression (35) is then substituted in (34) and all the terms are averaged over a stellar pulsation period in the sense of equation (8). This leads to the following equation :

$$\frac{1}{P} \ln\left(\frac{\mathcal{A}_0(t+P)}{\mathcal{A}_0(t)}\right) + \frac{V_{sw}}{L_{\text{eff}}}\left[-1 + \frac{r - r_0}{L_{\text{eff}}} \frac{\partial L_{\text{eff}}}{\partial r}\right] = \\ -\frac{V_{sw}}{L} + \frac{\ln(T_A)}{P} \quad (36)$$

If it is assumed that $\mathcal{A}$ is a strictly periodic function :

$$\mathcal{A}_0(t+P) \equiv \mathcal{A}_0(t) \quad (37)$$

then the first term in (36) is identical to 0. It is useful to rewrite (36) using the following dimensionless quantities :

$$\begin{aligned} \lambda &\equiv L/R_* \\ \lambda_{\text{eff}} &\equiv L_{\text{eff}}/R_* \\ \xi &\equiv r/R_* \\ \xi_0 &\equiv r_0/R_* \end{aligned} \quad (38)$$

where $R_*$ is the stellar radius. The differential equation (36) then becomes :

$$\frac{1}{\lambda_{\text{eff}}} - \frac{\xi - \xi_0}{\lambda_{\text{eff}}^2} \frac{\partial \lambda_{\text{eff}}}{\partial \xi} = \frac{1}{\lambda} - \frac{R_* \ln(T_A)}{P V_{sw}} \quad (39)$$

Note that this equation does not apply at the sonic point where $V_{sw} = 0$. At this point the flow does not transport wave action density and the formulation of the weak damping used in equation (34) yields no damping. There is therefore apparently only an increase in wave action density due to the periodic impulsive increase due to passing shock waves. This is of course un-physical. Strong damping of non-linear sound waves is not taken into account here. The non-linear damping terms will limit the wave action density at the sonic point to a finite value.

Equation (39) is a form of a Bernouilli differential equation. The solution is :

$$\lambda_{\text{eff}} = \left[\frac{1}{\xi - \xi_0} \int_{\xi_0}^{\xi} \frac{1}{\lambda} - \frac{R_* \ln(T_A)}{P V_{sw}} d\xi'\right]^{-1} \quad (40)$$

The constant of integration is set equal to 0 because $\lambda_{\text{eff}}$ is assumed to be finite at $\xi = \xi_0$. The term between brackets in (40) can easily be recognized as an average of the integrand over the region between dimensionless radii $\xi_0$ and $\xi$. Note that it is allowed to have $\lambda$ be a function of radius. The following notations for these averages can be introduced :

$$\begin{aligned} \frac{1}{\langle \lambda \rangle} &\equiv \frac{1}{\xi - \xi_0} \int_{\xi_0}^{\xi} \frac{1}{\lambda} d\xi' \\ \frac{1}{\langle \lambda_* \rangle} &\equiv \frac{1}{\xi - \xi_0} \int_{\xi_0}^{\xi} \frac{R_* \ln(T_A)}{P V_{sw}} d\xi' \end{aligned} \quad (41)$$

With these definition the expression for $\lambda_{\text{eff}}$ becomes :

$$\lambda_{\text{eff}} = \frac{\langle \lambda_* \rangle \langle \lambda \rangle}{\langle \lambda_* \rangle - \langle \lambda \rangle} \quad (42)$$

If $\langle \lambda_* \rangle < 0$ (i.e. if $V_{sw} < 0$) the effective damping length is $0 < \lambda_{\text{eff}} < \langle \lambda \rangle$. This is the situation in which the sound waves of type A can 'drain' back into the star. If $0 < \langle \lambda_* \rangle < \langle \lambda \rangle$ the effective damping length is $< 0$. This means that the wave action density locally increases exponentially with radius, rather than decreases. For all $\langle \lambda_* \rangle > \langle \lambda \rangle$ the effective damping length $\lambda_{\text{eff}}$ is larger than the 'intrinsic' damping length $\langle \lambda \rangle$.

$\langle \lambda_* \rangle$ is not constant as a function of radius, but a rough estimate of the order of magnitude of $\langle \lambda_* \rangle$ can be obtained by substituting typical values for cool giants in equation (41) :

$$\langle \lambda_* \rangle \sim 0.124 \frac{(P/100 \text{ d})(V_{sw}/1 \text{ km/s})}{(R_*/100 \text{ R}_\odot) \ln(T_A)} \quad (43)$$

Using again the results from the 'standard model' of Bowen (1988) with $P = 350$ d, $R_* = 270$ R$_\odot$, and an estimated $V_{sw} \approx 0.1$ km/s and $\ln T_A = 1$, the value of $\langle \lambda_* \rangle \approx 0.02$. This is roughly equal to the static density scale height of Bowen's 'standard model'.

Equating the two different length scales yields the maximum effective damping length. As far as the author is aware no work has been done on the damping of acoustic waves in the atmospheres and winds of cool giant stars. There is some work available on the excitation and damping of sound waves in homogeneous radiative media (cf. Hearn, 1972 ; Mihalas & Weibel Mihalas, 1983). Using the results from that work yields a rough estimate of the intrinsic damping length of between 10 and 1000 times the intrinsic wave length of the wave. Using these results, waves with an intrinsic wavelength in the range of 0.1 to 0.001 times the static density scale height are most likely to have an effective damping length approaching infinity. This corresponds with intrinsic wave lengths $\Lambda$ and periods $\tau$ of the acoustic waves of :

$$\Lambda \sim 4\,10^6 \text{ m} - 4\,10^9 \text{ m}$$
$$\tau \sim 15 \text{ min} - 1 \text{ d} \qquad (44)$$

Note that all of these estimates are quite uncertain since they rely on analyses of wave damping that are not really appropriate and simulations of shocks in the winds of cool giants with an incomplete treatment of the radiative transfer.

The original models of Pijpers & Habing (1989) for the winds of AGB stars match the observed properties remarkably well, but for quite large values of the damping length $\lambda$. Although this has been considered as a weakness of those models, it may be that such large values arise precisely because of the interaction between the pulsation driven shocks and the sound waves.

### 4.2. The gradient of the wave pressure tensor

The presence of sound waves in a stellar wind is felt through the gradient of the pressure tensor and not through the pressure tensor itself. One should be careful when calculating the average value of this gradient over a pulsation period. Gradient operators and the integral of the averaging operation do not necessarily commute because of the presence in the wind of the discontinuities from the pulsation. Koninx & Pijpers (1992) give an expression for the gradient of the non-linear wave pressure tensor for sound waves with a purely radial wave vector :

$$(\nabla \cdot \mathbf{P})_r = -c_s^2 \widetilde{\rho} I(1, \gamma+1) \left( \frac{2}{r} \frac{1 - I(0, \gamma) - I(1, \gamma)}{\gamma I(1, \gamma+1)} \right.$$
$$\left. + \frac{1}{u} \frac{du}{dr} \frac{1 - I(0, \gamma)}{\gamma I(1, \gamma+1)} - \frac{1}{a|\mathbf{k}|} \frac{da|\mathbf{k}|}{dr} \right) \qquad (45)$$

The force from the wave pressure is outward on average if the average over a pulsation period of expression (45) is larger than 0. The functions $I$ depend on the wave profile function $f$ and are defined by (Koninx & Pijpers, 1992) :

$$I(n, \gamma) \equiv \frac{1}{2\pi} \int_0^{2\pi} d\theta \frac{(a|\mathbf{k}|f')^n}{(1 + a|\mathbf{k}|f')^\gamma} \qquad (46)$$

The integral $I(1, \gamma+1)$ is less than zero for realistic profile functions $f$, such as a saw tooth profile and a sinusoidal profile, and for all values of $a|\mathbf{k}|$.

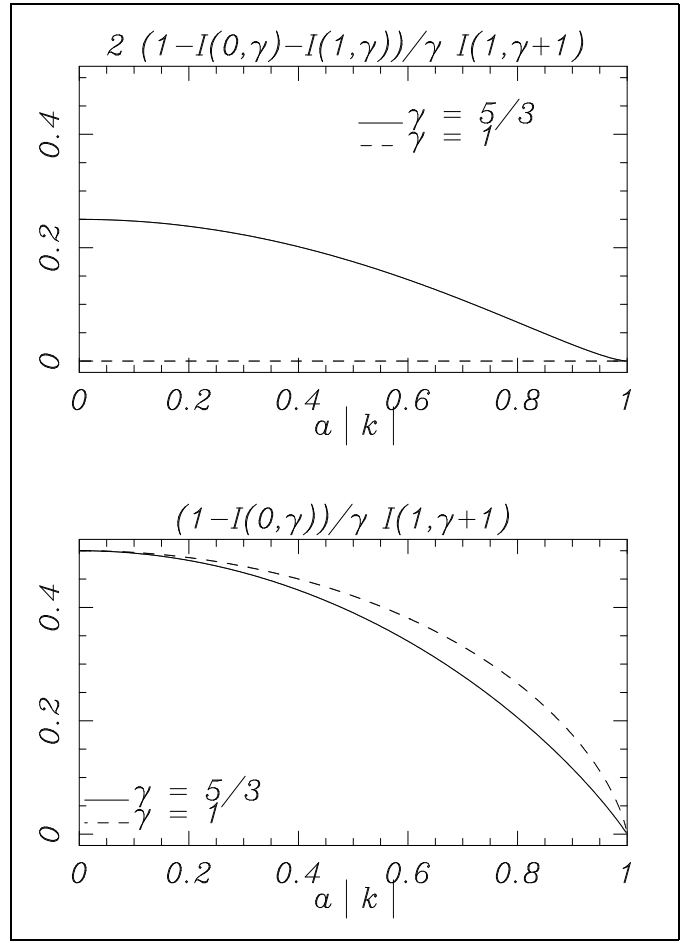

**Fig. 7.** The two terms involving the profile function integrals in equation (45) as a function of wave amplitude $a|\mathbf{k}|$. This is for a saw tooth profile function $f$.

The two terms involving the integrals $I$ for the geometrical factor $1/r$ and the velocity gradient term of equation (45) respectively are shown in figure 7 for a saw tooth profile $f$ and for two values of the adiabatic exponent $\gamma$. The same factors for a sinusoidal profile are not shown because they are virtually identical. It is clear that these terms are not negative for any value of the wave amplitude and therefore will also not be negative after averaging. The force from the sound waves can therefore point inward only if the average over a pulsation period of the logarithmic gradient of the wave amplitude is positive and larger than the first two terms in equation (45).

Although the waves damp as they propagate in the direction downstream from the shock this does not necessarily mean that the gradient of the amplitude $a|\mathbf{k}|$ must be positive everywhere. If the wind is exactly periodic, as considered in the previous section, a shock wave that passes must conserve the sign of this gradient. If a region in the wind has a negative gradient of the sound wave amplitude this gradient thus remains negative. The amplification of the sound waves by the shock is proportional to the sound wave action density. The constant of proportionality depends only on the amplitude of the pulsational shock. So if this shock has a constant amplitude over a certain region in the wind not only the sign but the also the value of the logarithmic gradient of the sound wave amplitude that appears in equation (45) remains unaffected by the passage of the shock.

From the analysis of the previous section it is clear that the wave action density and therefore the wave amplitude can increase quite rapidly with radius if $\langle\lambda\rangle > \langle\lambda_*\rangle$. It is clear from energy conservation arguments that the amplitude must decrease to zero towards infinity. Physically the damping length must become shorter as the wave amplitude increases so that at some point the effective damping length $\lambda_{\rm eff} > 0$. In principle the force from the sound wave pressure can therefore point inward on average in a part of the wind but it cannot do so throughout the wind. This means that the presence of shocks in the winds of Mira variables does not preclude that sound waves enhance the mass loss and/or the outflow velocity of the wind.

The relative importance of the sound waves and the pulsationally generated shocks in generating the mass loss is hard to assess on the basis of this work alone. In one of the models of Bowen (1992) the mechanical power delivered at the piston for a 5 $M_\odot$ model for driving at the fundamental mode period is $\sim 10\ L_\odot$ which amounts to a fraction of the radiative luminosity of $\sim 2.4\ 10^{-4}$. Pijpers and Habing (1989) show some results for a sound wave driven wind for a star with the same hydrostatic density scale height as the model of Bowen (Pijpers & Habing model 7). For sound wave driven winds this is the only relevant parameter, so the wind of each of these models can be compared. Bowen (1992) reports a quite low mass loss rate of $< 10^{-8}\ M_\odot$/yr whereas Pijpers and Habing (1989) show mass loss rates of up to $10^{-6}\ M_\odot$/yr for a mechanical luminosity in sound waves of $\sim 3\ 10^{-5}$ times the radiative luminosity. From this it would seem that the sound wave pressure alone is more efficient at generating mass loss than pulsational shocks alone for at least some stars. However this impression may be false for two reasons.

Firstly : because of the non-linear interaction between sound waves and periodic strong shocks described in this paper it is not at all clear what the situation is if both are present. Secondly : the velocity amplitude of the sound waves at the base of the wind is $\sim$ 10% of the sound speed for these models. This might appear to put the sound wave pressure in the regime where the linearized approach of Pijpers & Habing is still valid. However, in the wind the wave action density increases considerably, and therefore so does the amplitude, before decreasing to zero again. The amplitudes become so large that non-linear effects are important. As pointed out by Koninx & Pijpers (1992) this means that the results of the calculations done entirely in the linearized approach may not be reliable.

From figure 2 of Pijpers & Koninx (1992) it is seen that at larger amplitudes the sound waves generate a larger force than would be expected from an extrapolation of the linearized formulation. If this effect is primarily of importance in the region of the wind beyond the critical point this has the consequence that for the same final velocity in the wind, the mass loss rate is lower than predicted in the linearized models. If this effect primarily plays a role below the critical point, the mass loss is effectively increased. In the absence of actual modelling of non-linear sound wave driven winds, with and without periodic shocks generated by pulsation, it is premature to claim that either of these mechanisms dominate in the determination of the mass loss of giant and supergiant stars.

## 5. Implications for numerical modelling

The above results are important not only for the models of sound wave driven winds. The results of the numerical modelling of pulsation driven mass loss of Bowen (1988) and Wood (1979) show that pulsation can drive a mass loss only in combination with the radiation pressure on dust formed in the stellar wind. However the effect of sound waves is not taken into account in their models. It is possible that the combination of propagating shocks generated by the pulsation and sound waves generated by the surface convection zone can drive the mass loss of red giants even without the driving of radiative forces on dust.

It is important to realize at this point that sound waves are suppressed in **all** numerical hydrodynamical codes for reasons of numerical stability and also because of the finite spacing of grid points. All explicit hydrodynamical codes use artificial viscosity to stabilize shocks. If it is not used numerical inaccuracies will shift the shock back and forth over distances small compared to the grid spacing which causes the shock to quickly disperse. This is exactly the same effect as described in this paper but instead of actual sound waves perturbing the shock the sound is produced artificially by 'numerical jiggling' of the shock front. A high artificial viscosity is introduced to damp these oscillations and thereby prevent the dispersal of shocks. The effect of this high viscosity is not only to damp the numerical production of sound but it also suppresses real sound waves. The damping length $\lambda$ is artificially shortened and the dynamical effects of sound waves are destroyed. The usual recipe for artificial viscosity (Richtmyer & Morton, 1967) uses a coefficient viscosity $\eta$ of the form :

$$\eta = l^2 \rho |\nabla \cdot \mathbf{v}| \qquad (47)$$

Here $l$ is a numerical smoothing length. The Reynolds' number of sound waves with this artificial viscosity is :

$$\mathrm{Re} = \frac{\rho \Lambda \delta v}{\eta} = \frac{\Lambda \delta v}{l^2 |\nabla \cdot \mathbf{v}|} \qquad (48)$$

The limiting wave length of sound waves below which the viscosity dominates the dynamics can be estimated roughly by setting Re = 1. Away from shocks the local velocity divergence is probably of the order of the velocity amplitude $\delta v$ of the sound waves. Using this the minimum wavelength that can be represented is of the order of the numerical smoothing length $l$. At the shocks the divergence of the velocity can be orders of magnitude larger than the velocity amplitude of the sound waves. This is precisely the reason why artificial viscosity is introduced.

However the lower limit for wave length of sound waves that can still be properly represented increases considerably :

$$\Lambda_{\min} \gtrsim 10^2 l \tag{49}$$

All numerical hydrodynamical codes that employ artificial viscosity in some form can therefore not model winds in which sound waves play a role. This essentially leaves implicit or high resolution shock capturing (HRSC) numerical schemes as the only option, since these do not need artificial viscosity to ensure the numerical stability of shocks (see e.g. van Leer, 1979 ; Korevaar & van Leer, 1988).

For HRSC hydrodynamical codes there are still the problems of sampling in time and sampling in space. Although the time step is not limited by the CFL condition, it is limited because the sound wave is only well sampled for $n_t$ points per period.

$$\Delta t \leq \frac{P_{\rm sw}}{n_{\rm t}} \tag{50}$$

Here $P_{\rm sw}$ is the minimum period of the sound waves to be modelled. Because the period of the sound waves can be quite short compared to the period of the shock waves, it would take many time steps to calculate wind models.

The spatial sampling problem is much more serious. The effects of discrete spatial sampling of the flow are sometimes also referred to as *numerical* viscosity. All such effects are an unavoidable consequence of the discretization of the fluid equations as opposed to *artificial* viscosity which is added as an explicit term in these equations. The finite spacing of grid points causes a sampling of the sound waves that is less accurate as the wavelength of the sound wave decreases. If the sound wave is considered to be sufficiently well sampled for a minimum of $n_x$ points per wavelength the maximum grid spacing for a given period of the sound waves that are to be modelled is :

$$\Delta x \leq \frac{P_{\rm sw} \min(v_{\rm sw})}{n_{\rm x}} \tag{51}$$

where $\Delta x$ is the spacing of the numerical grid and $\min(v_{\rm sw})$ is the minimum of the sound wave propagation speed in the rest frame of the star. Since the wind is likely to have at least one transition from subsonic to supersonic outflow at any time, this means that near that point the wavelength in the Eulerian frame decreases to 0. This implies that the required grid spacing would also decrease to zero. Thus the required resolution is always higher than can be achieved with a finite number of grid points. Therefore even an HRSC type adaptive grid method can not properly conserve wave action.

All schemes that solve the fluid equations on a discrete mesh in space and/or time suffer from this effect of numerical viscosity. None of these schemes can therefore properly conserve wave action. The **only** solution is to include an equation like (34) into the scheme of equations to be discretized and solved simultaneously with the equations of conservation of mass, momentum, and energy.

## 6. Conclusions

In this paper several points are made. The first point is that in stellar winds the interaction of sound waves with strong shocks generated by pulsation does not remove wave action density of the sound waves. It is shown that the sound wave action density increases upon the passage of a shock. This is relevant to the models of the winds of Mira variable stars and other pulsating stars that lose mass through a stellar wind.

The effective damping length of the sound waves in the rest frame of the star can increase considerably because of the periodic injections of wave action density through the passage of shocks. The combination of pulsation driven shocks and sound waves could therefore justify the large values of the dimensionless damping length used by Pijpers and Habing (1989). However, the functional form of the wave action density differs from the one used by these authors and it is necessary to do similar work for the combination of sound waves and pulsation before the effectivity of this model can be discussed.

The time dependent hydrodynamical simulations of the winds of Miras carried out to date do not and cannot properly take into account the dynamical effect of the presence of sound waves. A proper treatment requires the solution of the transport equation for wave action density jointly with the equations for mass, momentum and energy. If sound waves are generated in the surface convective layers of cool giants in much the same way as they are in the convective layer of the sun, their dynamical effect on the winds of cool giants should not be ignored. This means that the paradigm for winds of AGB stars, exemplified by the work of Wood (1979), Bowen (1988), and Bowen & Willson (1991), needs to be adjusted.

*Acknowledgements*. The author is grateful to the referees J. Dyson and R. Williams for their questions and comments which helped to improve this paper.